\documentclass[9pt,preprint2]{aastex}
\usepackage{amssymb,graphics,latexsym,amsmath}
\usepackage{rotating}

\newcommand{\eighteen}{PSR J1853+1303}
\newcommand{\nineteenofive}{PSR J1905+0400}
\newcommand{\nineteenten}{PSR J1910+1256}
\newcommand{\nineteenfiftythree}{PSR B1953+29}
\newcommand{\twenty}{PSR J2016+1948}
\newcommand{\msun}{\mbox{M$_\odot$}}

\begin{document}
\title{High-Precision Timing of 5 Millisecond Pulsars: Space Velocities, Binary Evolution and Equivalence Principles}
\author{M. E. Gonzalez\altaffilmark{1}, I. H. Stairs\altaffilmark{1}, R. D. Ferdman\altaffilmark{2}, P. C. C. Freire\altaffilmark{3}, D. J. Nice\altaffilmark{4}, 
P. B. Demorest\altaffilmark{5}, S. M. Ransom\altaffilmark{5}, M. Kramer\altaffilmark{3}, F. Camilo\altaffilmark{6}, G. Hobbs\altaffilmark{7}, 
R. N. Manchester\altaffilmark{7}, A. G. Lyne\altaffilmark{2}}

\altaffiltext{1}{Department of Physics and Astronomy, University of British Columbia, 6224 
Agricultural Road, Vancouver, BC, V6T 1Z1, Canada; gonzalez@phas.ubc.ca1}
\altaffiltext{2}{Jodrell Bank Centre for Astrophysics, University of Manchester, Manchester, M13 9PL, UK}
\altaffiltext{3}{Max-Planck-Institut f\"{u}r Radioastronomie, Auf dem H\"{u}gel 69, D-53121 Bonn, Germany}
\altaffiltext{4}{Physics Department, Lafayette College, Easton, PA 18042, USA}
\altaffiltext{5}{National Radio Astronomy Observatory, Charlottesville, VA 22903, USA}
\altaffiltext{6}{Columbia Astrophysics Laboratory, Columbia University, New York, NY 10027, USA}
\altaffiltext{7}{Australia Telescope National Facility, CSIRO, Epping NSW 1710, Australia}

\begin{abstract}
We present high-precision timing of five millisecond pulsars (MSPs) carried out for more than seven years; four 
pulsars are in binary systems and one is isolated. We are able to 
measure the pulsars' proper motions and derive an estimate for their space velocities. The measured 
two-dimensional velocities are in the range 70--210 km~s$^{-1}$, consistent with those measured for other MSPs. 
We also use all the available proper motion information for isolated and binary MSPs to update the known
velocity distribution for these populations. As found by earlier works, we find that the velocity distribution of 
binary and isolated MSPs are indistinguishable with the current data. 
Four of the pulsars in our observing program are highly recycled with low-mass 
white dwarf companions and we are able to derive accurate binary parameters for these systems. For three of 
these binary systems we are able to place initial constraints on the pulsar masses with best-fit values in the 
range 1.0--1.6~\msun. 
The implications of the results presented here to our understanding of binary pulsar evolution are discussed. 
The updated parameters for the binary systems studied here, together with recently discovered similar 
systems, allowed us to update previous limits on the the violation of the strong equivalence principle through 
the parameter $|\Delta|$ to 4.6$\times$10$^{-3}$  (95\% confidence) and the violation of 
Lorentz-invariance/momentum-conservation through the parameter $|\hat{\alpha}_3|$ to 5.5$\times$10$^{-20}$ 
(95\% confidence). 

\end{abstract}


\section{Introduction}
Pulsars are believed to be born with spin periods of $\sim$0.1~s and gradually slow down as they 
age due to the loss of rotational kinetic energy in the form of electromagnetic radiation. These ``normal" pulsars make 
up the bulk of the observed population and $\sim$1,900 of them are currently known \citep{mhth05}\footnote{See
also the ATNF Pulsar Catalogue: \url{http://www.atnf.csiro.au/research/pulsar/psrcat/}.}.
On the other hand, a group of old, fast-spinning pulsars is observed ($\sim$200 pulsars). It is believed that these 
old pulsars are formed from the transfer of mass and angular momentum 
from a previous or present companion in a binary system \citep{acrs82}. These 
pulsars are generally seen as ``recycled'' members of the population: old pulsars that have been spun-up and 
brought back to an active, pulse-emitting life thanks to their interaction within a binary system. The fastest spinning
pulsars known have spin periods of $P_s$~$\lesssim$~0.01 s, so-called ``millisecond pulsars'' (MSPs), and are thought 
to have been produced in this manner. 
The measured characteristics of the members of these binary systems and their
orbital parameters provide valuable insights into the formation and evolution of these systems. 
See, e.g., Stairs (2004) and Lorimer (2008)\nocite{sta04a,lor08} for general reviews of binary pulsars and their
scientific importance.

The measurement of a pulsar's proper motion can be used to estimate its space velocity 
\citep[e.g.,][]{hllk05,cbv+09}. Such measurements are 
important for a variety of scientific questions, including estimating the distribution of natal kicks imparted to 
proto-neutron stars by the supernova (SN) explosion that created them. A variety of mechanisms have been proposed
that can give rise to these natal kicks \citep[e.g.,][]{sp98,ks99b,jsk+05}. Another key question is whether isolated MSPs have 
a similar velocity distribution to those still in binary systems \citep[e.g.,][]{tb96,tsb+99,lkn+06,hllk05,mlc+04}. In general, 
we expect MSPs to have lower system velocities than the 
rest of the pulsar population since, after the SN explosion, the binary system must have remained intact to spin up the
neutron star. However, for isolated MSPs, the companion must eventually leave the system or be evaporated. 
Arguments for both lower and higher velocities for isolated MSPs have been given in the literature \cite[e.g.,][]{tsb+99,mlc+04}.
Given that only a small number of isolated MSPs with measured proper motions are known ($<$10 objects), additional 
measurements are very important.

Many subclasses of pulsar binary systems are now recognized \citep[e.g.,][]{sta04a,lor08}. The broader distinction 
made is between those pulsars with 
high-mass companions (e.g., another neutron star) and those with lower mass companions (e.g., white dwarfs -- 
WDs). In the case of pulsars with WD companions, various subgroups are generally identified. For example, 
mildly-recycled pulsars ($P_s$ $\sim$ tens of milliseconds) in a tight orbit (orbital periods $P_b$ $\lesssim$ a few 
days) with high-mass WD companions  ($m_2$~$\sim$~1~\msun) are thought to arise from a common-envelope 
evolution or from periods of ultra-high mass transfer during a Roche-lobe overflow phase \cite[][]{vdh94a,tvs00}. 

A more straightforward evolution is thought to apply for MSPs in long orbits ($P_b$~$\gtrsim$~4 days) with 
low-mass WD companions ($m_2$~$\lesssim$~0.3~\msun), generally called wide-orbit binary millisecond pulsars
\citep[WBMSP;][]{rpj+95,ts99}. Here, as the companion evolves and overflows its
Roche-lobe during the red-giant phase, mass spirals onto the neutron star and forms an accretion disk. A stable,
long-lived mass transfer phase is expected to take place, producing nearly circular orbits (eccentricities of 
$e$~$\sim$~10$^{-6}$--10$^{-3}$). \citet{phi92} predicted that these systems should exhibit an 
orbital period--eccentricity ($P_b - e$) relationship based on the expectation that convective eddies in the
envelope of the red giant will produce nonzero values of the eccentricity. The point at which mass transfer
ceases and $e$ freezes depends on the size of the red giant envelope, which will in turn determine the size
of the orbit and thus the orbital period. In addition, the size of the envelope is also thought to be related to the
mass of the red giant's core, which eventually contracts to form a WD. Therefore, an orbital period--core mass
($P_b-m_2$) relationship is also expected in these systems \citep{rpj+95,ts99}.

The WBMSP systems also provide important tests for theories of gravity. For example, 
the strong equivalence principle (SEP) states that all neutral test masses 
fall with the same acceleration in an external gravitation field, i.e., it states that the gravitational and 
inertial masses of self-gravitating bodies are identical ($m_g$/$m_i$~$\equiv$~1). Binary pulsar systems allow 
us to test for SEP violations in the limit of high self-gravity \citep{ds91,wex97,wex00,sfl+05}: if the binary 
components experience different accelerations in the
gravitational field of the Galaxy, a forced eccentricity is imparted to the system along the projected direction
of the external force onto the orbital plane. Binary pulsars with low companion mass, small eccentricity and
long orbital periods are ideal for SEP violation tests.

Another important test of gravitational theories involves the 
post-Newtonian parameter $\alpha_3$ which is associated with the violation of momentum conservation and the
existence of preferred frames \citep[Lorentz invariance;][]{wil93}. In general relativity (GR), $\alpha_3$~$\equiv$~0. 
The most observable
effect of a possible deviation from this GR prediction is thought be a non-zero self-acceleration for a rotating body in a
direction perpendicular to its spin axis and perpendicular to its velocity with respect to the absolute rest frame 
\citep{bd96}. 
In the case of binary systems, each component will experience self-acceleration. These self-accelerations 
perturb the orbital dynamics, leading to a forced eccentricity and polarization of the orbit along a fixed direction.

Here we present results obtained from long-term timing of 5 MSPs, four of which are WBMSPs (\eighteen, 
\nineteenten, \nineteenfiftythree, \twenty)  and one of which is isolated (\nineteenofive). We use our results to 
study the space velocities of MSPs, binary evolution models and equivalence
principle tests. In \S\ref{sec:data} we describe the observations performed and data analysis carried out. 
In \S\ref{sec:results} we present our improved timing solution for these pulsars, including the measured proper
motions. 
In \S\ref{sec:vels} we 
discuss the implications for the velocity distribution of isolated and binary MSPs and discuss their implications
for binary evolution models of MSPs. 
In \S\ref{sec:xdot} we derive constraints on the component masses for three systems and we discuss these results 
in light of evolution models. In \S\ref{sec:sep} we present updated upper limits of equivalence
principle violations using WBMSPs. Finally, in \S\ref{sec:conclusion} we summarize our findings and point to future 
directions of this research.

\section{Observations and Data Analysis}\label{sec:data}
We have conducted high-precision timing on four WBMSPs and one isolated MSP. We collected data from two 
observatories and a total of five data acquisition systems. 
Here we describe the observing setups used at each telescope. A summary of the observations for each pulsar
is given in Table \ref{tab:obs}.

\subsection{Arecibo}\label{ssec:ao}
All pulsars were observed with the 305-m Arecibo telescope in Puerto Rico. The Wideband Arecibo Pulsar
Processors \citep[WAPPs;][]{dsh00} were used to observe all the pulsars. Three of the four WAPPs were used for most 
pulsars, except \twenty\ for which all four WAPPs were used in some observations. They were operated in online folding 
mode with 32~$\mu$s  sampling and 192 lags near 1400 MHz and 96 lags at 2700 MHz (for \twenty\ a sampling time of
128~$\mu$s and 128 lags were used). The Arecibo Signal Processor \citep[ASP;][]{dem07} was used for all pulsars except 
\twenty. ASP provides 0.25$\mu$s complex sampling in two orthogonal polarizations. These data were coherently de-dispersed 
in software using 16 or 24 frequency channels, each with 4 MHz bandwidth. The polarisations were later summed and the 
signal folded at the pulsar period. The ASP observations
were flux calibrated with a pulsed noise diode of known strength and, when available, with observations of a standard 
flux calibration source. A typical observing session involved collecting multiple integrations of 1 or 3 min long on each 
pulsar with ASP and the WAPPs for a total of up to 30 min of data. The data from the short integrations were aligned and summed 
using a preliminary timing model. For the WAPPs, all data were summed to obtain at single profile for each observation. 
For ASP, a separate profile was obtained at each frequency channel for each observation.

For \twenty\ we also used data from the Penn State Pulsar Machine \citep[PSPM;][]{cad97}, an analogue 
filterbank with 128 $\times$ 60 kHz frequency channels. The power level for each 128 channels was
sampled every 80~$\mu$s and stored to tape. The data were subsequently folded and aligned multiple times
as the ephemeris for the pulsar was being refined. The PSPM and WAPP data for this pulsar were
summed every few minutes and multiple profiles were obtained for most epochs.

To provide a long time baseline and improve the measured proper motion
for \nineteenfiftythree\ we also used data taken with the Mark II system \citep{raw86}, a dual-polarization 
32 $\times$ 30 kHz filterbank spectrometer. Here, the outputs from opposite polarizations were summed and
sampled at the pulsar period. Data were averaged over intervals of 1-2 minutes and stored for off-line processing.
A total of around 1 hour of data were collected at each epoch and the data summed to obtain one profile
for each observation\footnote{Additional data for \nineteenfiftythree\ from Arecibo (MJD 49129--49255)
and the Effelsberg telescope in Germany (MJD 49768--50460) were also available but did not add significantly to the 
results and were left out of our analysis.}. 

\subsection{Parkes}\label{ssec:pks}
We used the 64-m Parkes telescope in NSW, Australia to observe pulsars \eighteen, \nineteenofive\ and 
\nineteenten. These observations were taken using the Parkes analogue filterbank centered at 1390 MHz with 
512 $\times$ 0.5-MHz frequency channels sampled every 0.25 ms. Two polarizations were recorded and summed
in hardware for each frequency channel. The data were subsequently folded off-line using a preliminary ephemeris
and summed to obtain a single profile for each observation.\\


A time of arrival (TOA) was found for each observation by cross-correlating the profiles with a high signal-to-noise 
standard template \citep{tay92}. The recorded observatory times were corrected to UTC time by using data from 
GPS satellites. The JPL DE405 ephemeris \citep{sta04b} was used for barycentric corrections.
The software package TEMPO\footnote{http://tempo.sourceforge.net/.} was used to find the timing solution for
each pulsar by including astrometric, binary and spin parameters as needed to arrive at a phase-connected
solution (where every rotation of the star over the span of the observations is accounted for). In order to fit
for any instrumental or standard template profile differences, we fit for arbitrary time offsets between each 
instrument (for ASP we have also allowed for time offsets between each 4~MHz channel to fit for any profile
changes across its bandwidth). A change in dispersion measure (DM) over time was detected for \twenty\ and 
marginally for \nineteenofive\ and \nineteenfiftythree\ (see Table \ref{tab:pars}). 
Finally, the measured uncertainties were scaled by a small telescope-dependent 
amount to ensure a timing fit with $\chi_\nu^2\simeq$1.  

\section{Results and Discussion}\label{sec:results}
\begin{figure}
\begin{center}
\includegraphics[width=15cm]{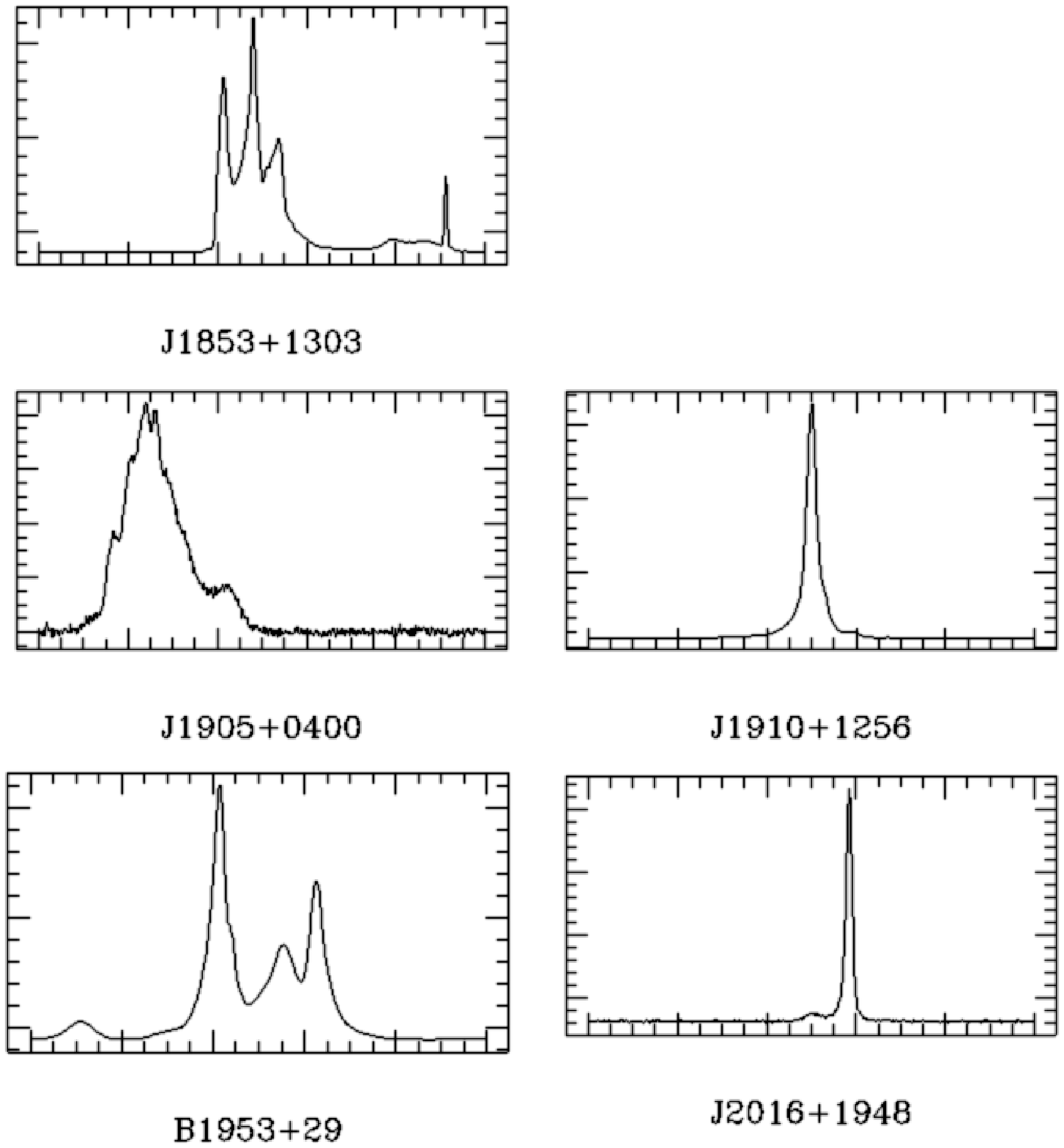}
\caption{\label{fig:std} Standard profiles for each pulsar at 1400 MHz obtained by combining all the Arecibo data used in our analysis. 
For \twenty\ we used the WAPP data and for all other pulsars we used the ASP data. The x-axis shows one pulse period and the y-axis
shows arbitrary flux units.}
\end{center}
\end{figure}

\begin{figure}
\begin{center}
\includegraphics[width=15cm]{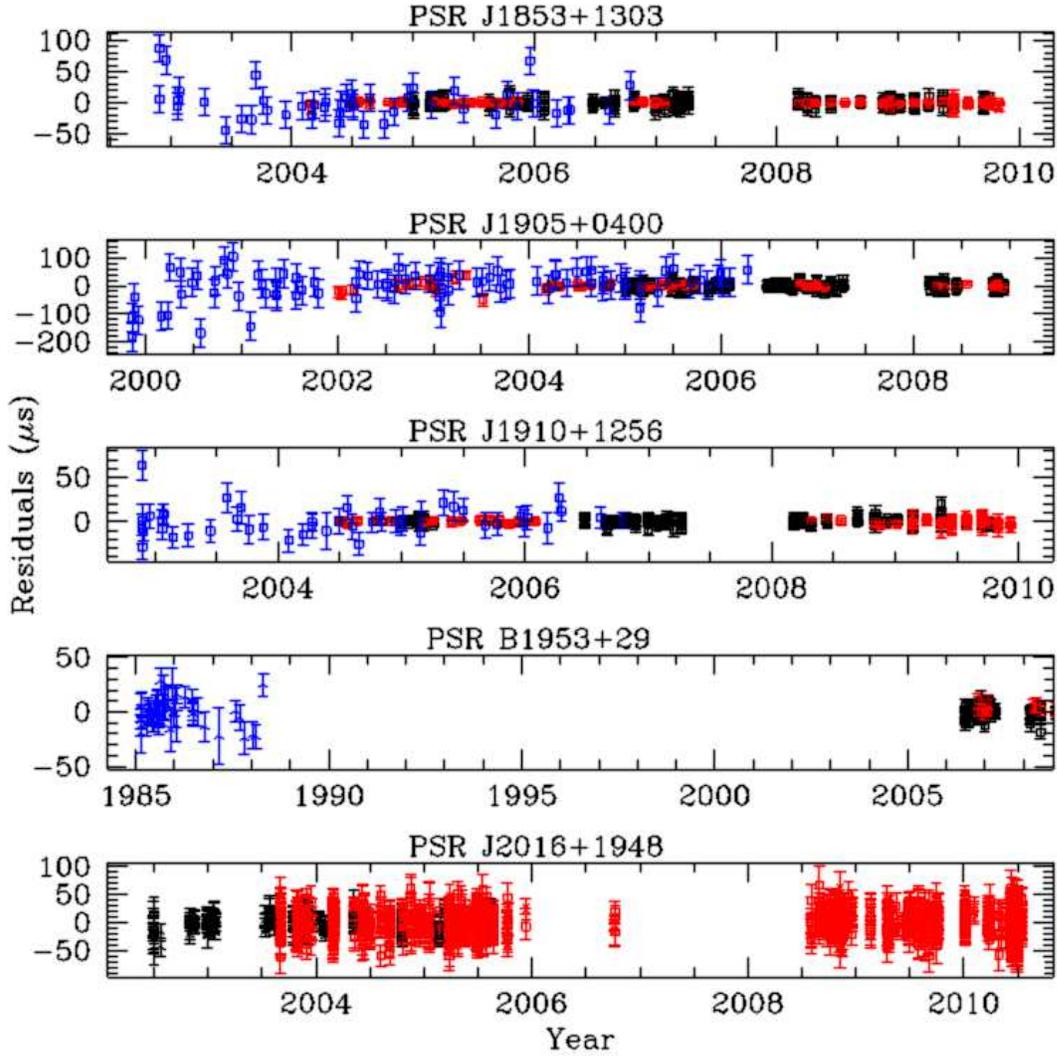}
\caption{\label{fig:res} Post-fit timing residuals for each pulsar. From top to bottom we show: \eighteen,
\nineteenofive, \nineteenten, \nineteenfiftythree\ and \twenty. For all pulsars, the black TOAs are those obtained
from ASP (except for \twenty, where black data show the TOAs obtained with PSPM). In all cases, the red 
TOAs are from the WAPPs. For all pulsars, the blue TOAs are those obtained from Parkes (except for \nineteenfiftythree,
where the blue TOAs are those from the Mark II instrument).}
\end{center}
\end{figure}

In Figure \ref{fig:std} we show the standard pulse profiles for each pulsar at 1400~MHz. Table \ref{tab:pars} shows the timing 
solutions derived from our work and Figure \ref{fig:res} shows the timing residuals derived from these solutions. The timing 
solutions successfully model the measured TOAs and leave no significant trends in the residuals. Pulsars \eighteen\ and 
\nineteenten\ have very low root-mean-square (rms) values and are now part of a long-term timing program to detect and study 
gravitational waves using an array of well-timed pulsars \citep{haa+10,dll+09,dgn+11}.

\eighteen\ and \nineteenten\ were discovered by the Parkes multibeam pulsar survey \citep[e.g.,][]{mlc+01} and their
timing solutions were first reported by \citet{sfl+05}. The longer data span possible with the Parkes data and the high 
quality of the Arecibo data allowed us to improve the timing solutions for these pulsars (especially for their binary 
parameters) and, for the first time, report a measurement of their proper motions. For \nineteenten, we are also able 
to measure a secular change in its projected semimajor axis (a similar, though marginal, measurement was also 
made for \eighteen). 

\nineteenfiftythree\ was discovered while performing a systematic search for radio pulsars using Arecibo 
in position error boxes from $\gamma$-ray sources reported by the COS B satellite \citep{bbf83}. Previous timing 
solutions for \nineteenfiftythree\ have been reported by \citet{rtd88} and \citet{wdk+00}. Here we have been able to use 
Arecibo data spanning 25 yrs to derive a much improved timing solution for this pulsar, and especially so for its proper 
motion measurement. \nineteenofive\ was also discovered by the Parkes multibeam pulsar survey \citep{hfs+04}
and is one of only $<$20 known isolated MSPs. Here we are able to measure for the first time a proper motion for
this pulsar.

\twenty\ was discovered in the Arecibo 430~MHz Intermediate Latitude Survey \citep{naf03}. The discovery dataset for 
this pulsar covered about a year (taken in 1999) and were enough to determine that the pulsar was in a binary 
system with an unusually long orbital period of 635~days. However, deriving a complete timing solution from this initial
dataset was not possible. It was later discovered that the original 1999 data have large systematics, 
most likely the result of being folded with an inacurate estimate for the spin period. These data have been left out of our 
analysis and all subsequent data were taken in observing modes that allow for re-folding and re-aligning as the pulsar 
ephemeris was being improved. The current timing solution leaves no systematics in the derived TOAs and has been 
correctly predicting new TOAs for many years. We are therefore confident that we have found the most  
precise timing solution currently available for this system. Our timing solution in Table \ref{tab:pars} shows that the 
pulsar is indeed recycled and in a nearly circular orbit, likely 
the result of mass transfer and tides as its companion was going through its giant phase.
We are also able to measure the proper motion for this system and a secular change in the projected semimajor axis. 

\begin{figure}
\begin{center}
\includegraphics[width=15cm]{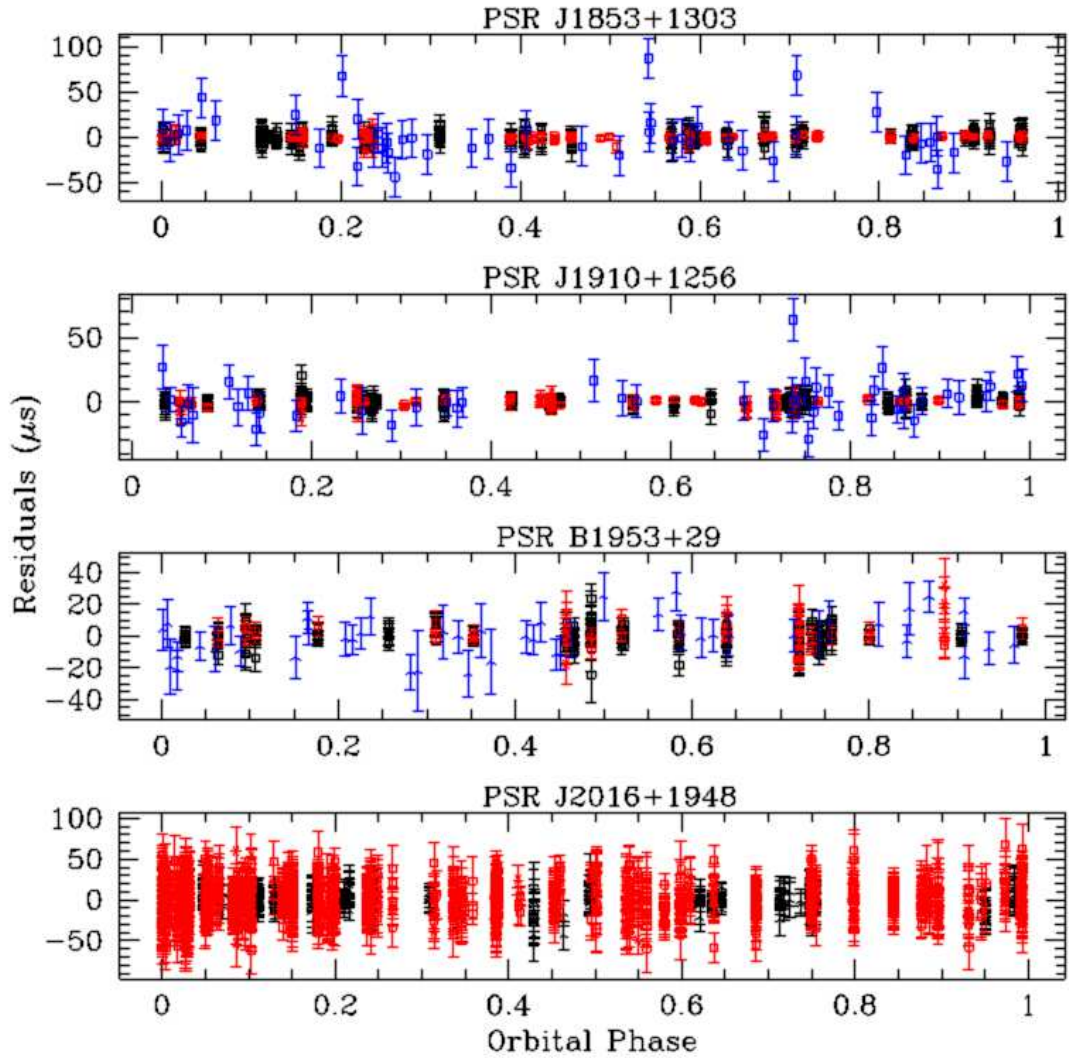}
\caption{\label{fig:resorb} Same as for Figure \ref{fig:res} but showing only the binary pulsars in our sample with 
their residuals plotted as a function of orbital phase.}
\end{center}
\end{figure}

None of the pulsars show a significant value of annual parallax (see Table \ref{tab:pars}). Only \eighteen\ has a 
marginal parallax measurement with a large error. It is possible that further observations of this pulsar with improved timing 
precision will be able to produce a more constraining parallax measurement. In addition, none of the binary pulsars 
show a measurable Shapiro delay. The residuals obtained from the best-fit timing solution are shown as a function of 
binary phase in Figure \ref{fig:resorb}. 

\subsection{Millisecond Pulsar Velocities}\label{sec:vels}
The high precision obtained by our timing study allowed us to measure a statistically significant value for the proper motions
of all five pulsars (see Table \ref{tab:pars}). We have used these measurements to study the velocity distribution of MSPs, 
both isolated and those in binary systems. The pulsar population in general has been found to have large space velocities
with a mean of $\sim$300~km~s$^{-1}$ \citep{ll94,hllk05}. Recycled MSPs appear to be on the low end of the 
velocity distribution, with a mean of $\sim$90~km~s$^{-1}$. In addition, no significant difference has been found between
the velocities distributions of isolated MSPs and those still in binary systems \citep{hllk05,lkn+06} despite the fact that an 
additional evolutionary stage (the disruption of the binary) has occurred in the former.

Now we revisit the velocity distribution of MSPs, which we take to be those with periods of $P$~$<$~0.01~sec and are
therefore fully recycled. From the new timing solutions presented in Table \ref{tab:pars}, only \twenty\ is not a fully 
recycled MSP and its implied 2D velocity of 96~km~s$^{-1}$ (at the implied DM distance of 2.5~kpc) was not included in 
the following analysis. \nineteenofive\ 
studied here is particularly important, as it is one of only ten isolated MSPs with measured proper motions.
Our sample then consists of 10 isolated MSPs and 27 binary MSPs for a total of 37 pulsars. We have combined the
measured proper motions with the available distance estimates to calculate the pulsars' 2D space velocities in their 
respective local standard of rest at the pulsar location. To do this we have corrected for Solar motion and used a peculiar velocity for the 
Sun of $V_\odot$=13.4~km~s$^{-1}$ \citep{db98}. We have also assumed a flat Galactic rotation curve with a 
galactocentric distance for the Sun of $R_\odot$=8 kpc and a Galactic rotation velocity of 222~km~s$^{-1}$
\citep{esg+03,db98}. A flat Galactic rotation curve is thought to be a good approximation for distances from the Galactic 
centre of $>$3~kpc (Olling \& Merrifield 1988; the pulsars we used have distances to the Galactic center of $>$4~kpc)\nocite{om98}. 
The resulting 2D velocities in the pulsars' standard of rest after correcting for Solar and Galactic motion, $V_{2D}$, are shown in Table 
\ref{tab:vels}.

Most pulsars 
have distance estimates from timing measurements of their dispersion measure (DM) combined with a model of the free 
electron distribution in the Galaxy \citep{cl02b}. In general, distances derived using this method are thought to have a 
$\sim$25\% error, implying a minimum similar error on the estimated velocities. For individual pulsars, the distance error could 
be much larger than that. The errors on the estimated pulsar velocities shown in Table \ref{tab:vels} were derived using Monte 
Carlo simulations with 10,000 runs per pulsar. For these simulations, pulsar distances were drawn from Gaussian distributions
centered on the values listed in Table \ref{tab:vels} with a width of 25\% of the central value\footnote{This 
uncertainty is consistent with the $\sim$20\% estimate of uncertainty in distances due to unmodeled inhomogeneities in the 
interstellar medium model \citep{cl03}.} (for pulsars where non-DM distances
are available the corresponding distance errors were used). Pulsar proper motions were then drawn using Gaussian distributions
with central values and widths derived using the values listed in Table \ref{tab:vels}. In practice, the largest error contribution to 
the estimated pulsar velocities are the associated distance errors. 
We then used the velocities in Table \ref{tab:vels} to study the velocity distribution of MSPs.

\begin{figure}
\begin{center}
\includegraphics[width=15cm]{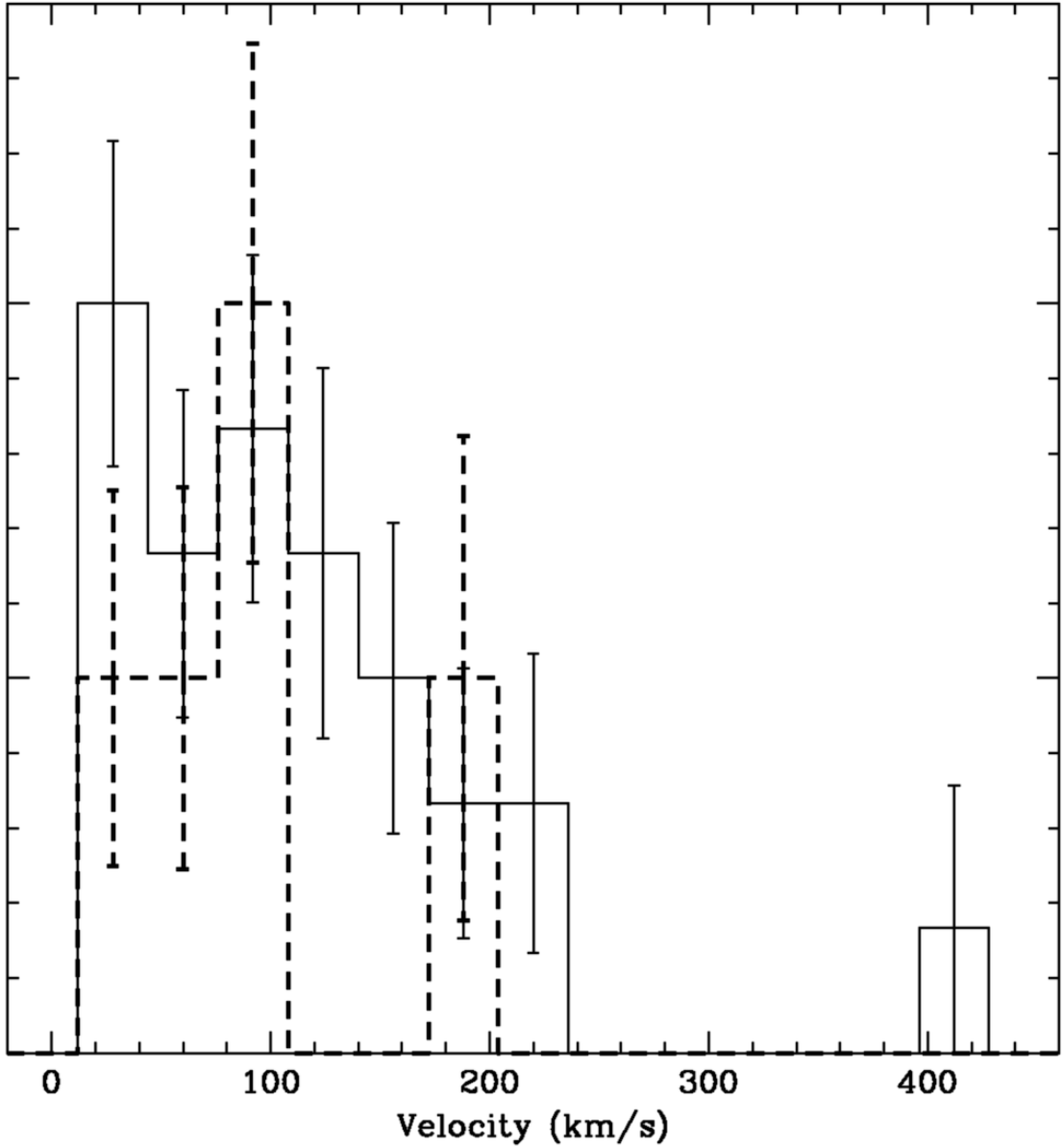}
\caption{\label{fig:vels} Normalized histograms of the 2D velocity distribution of binary (solid line) and isolated 
(dashed line) MSPs. The errors for each bin are given by the propagated measurement errors.}
\end{center}
\end{figure}

Figure \ref{fig:vels} shows the normalized histograms of the 2D velocities of binary (solid 
line) and isolated MSPs (bold dotted line) in our sample. The average velocities are found to be
108$\pm$15, 113$\pm$17 and 93$\pm$20 km~s$^{-1}$ for all MSPs,
binary MSPs and isolated MSPs, respectively\footnote{The errors in the average velocities shown in 
this section represent the standard errors of the mean.}. The average 2D velocities without correcting
to the pulsars' standard of rest are 88$\pm$12, 96$\pm$15 and 68$\pm$16 km~s$^{-1}$ for all, binary,
and isolated MSPs, respectively. The updated velocities are consistent with previous work: \citet{hllk05} 
reported 2D uncorrected velocities for binary and isolated MSPs of 
89$\pm$15 and 76$\pm$16~km~s$^{-1}$, respectively, with \citet{lkn+06} and \citet{mlc+04} reporting
similar values. We then find that the average velocities of binary and isolated MSPs are consistent with 
each other. To test whether the two samples are consistent with arising from the same distribution 
we use the Kolmogorov-Smirnov (KS) test \citep{mas51}\footnote{While the KS test doesn't take the error 
estimates into account, it is one of the most useful and general methods for comparing two samples. 
Detailed simulations to account for the errors are beyond the scope of this paper and can be carried 
out in future work.}. A KS test of the two corrected velocities results in a probability 
of 62\% that they are drawn from the same distribution. For the uncorrected 2D velocity measurements, the 
two distributions have a KS probability of 75\% that they are drawn from the same distribution. 

We therefore conclude that there is no statistically significant difference between the velocity distributions 
of isolated and binary MSPs with the current statistics. However, we also note that due to selection effects
our sample is biased towards nearby, low-velocity pulsars. It is therefore possible that the lack of difference 
between the velocity distribution of isolated and binary MSPs is due to our observing the low velocity tail 
of these distributions, which in reality could be quite different. Higher number statistics (particularly for
isolated MSPs) will allow for a more detailed study of such effects in the future. In addition, more distant
MSPs are now being discovered in current surveys (e.g., in addition to PSR J1903+0327 in Table \ref{tab:vels},
the velocities for two distant MSPs will also be published by Deneva et al. 2011\nocite{dcr+11}). 
While measuring the proper motions of distant objects will most likely require large amounts of telescope 
time, they represent significant additions to our sample.

\begin{figure}
\begin{center}
\includegraphics[width=15cm]{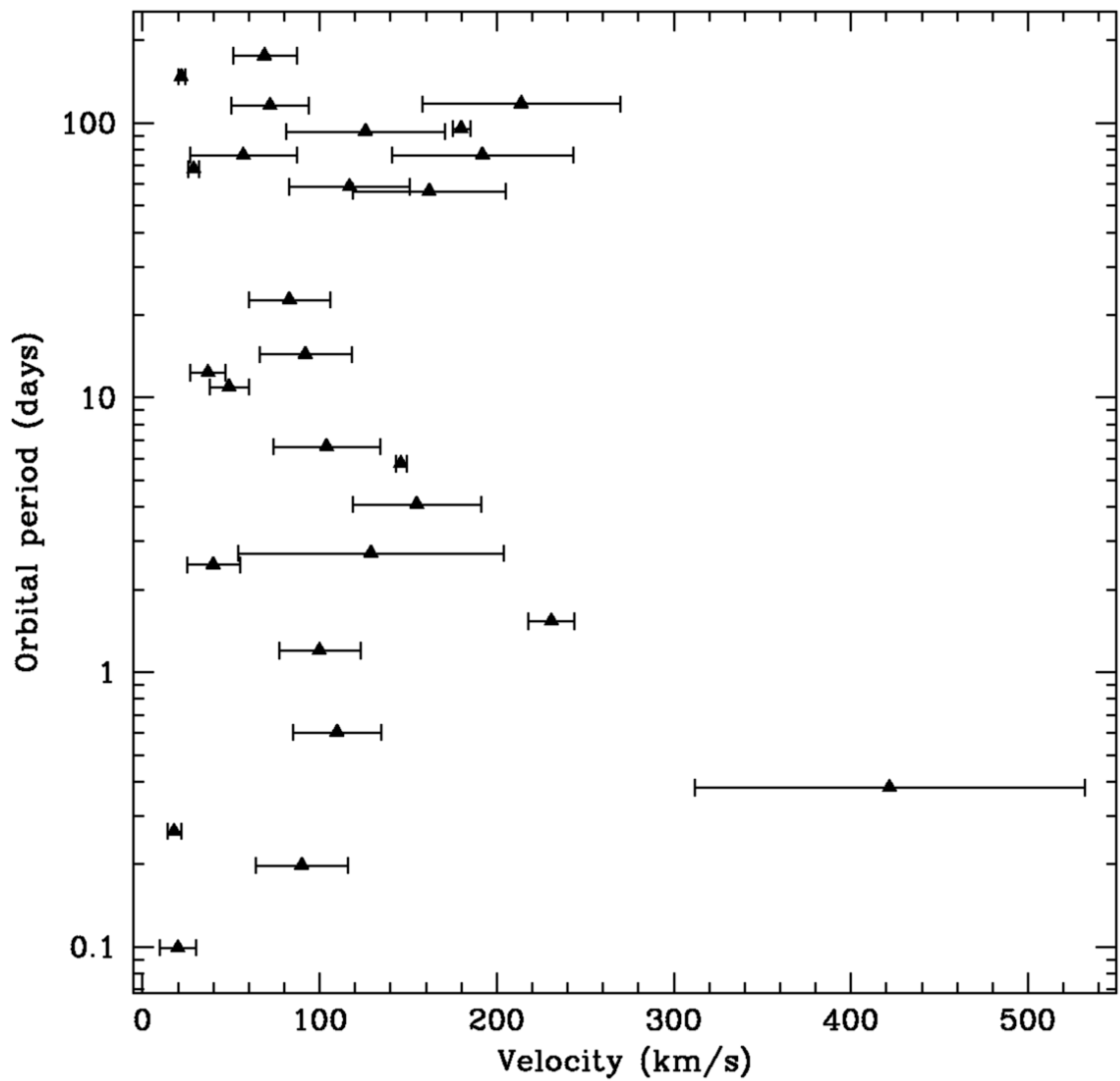}
\caption{\label{fig:velspb} Orbital period versus 2D velocities for binary MSPs. }
\end{center}
\end{figure}

Furthermore, we note that \citet{tb96} presented the expected space velocities of binary MSPs using various evolutionary
models. In their simulations, binaries with shorter periods tend to have larger velocities since the final velocity of the system
depends on the separation of the components at the time of the supernova explosion. However, this correlation is fairly weak 
and asymmetries in the explosion would easily wipe out this effect. \citet{hllk05} found no evidence for a correlation between 
the velocity of binary MSPs and their binary periods\footnote{\citet{hllk05} used a definition of $P<$~0.1~s and 
$\dot{P}<$~10$^{-17}$~s~s$^{-1}$ 
for recycled pulsars, thus including a mix of companion types in their sample. Here we use $P<$~0.01~s, resulting
in binaries with mostly helium white dwarf companions.}. We now briefly revisit this idea and in Figure \ref{fig:velspb} we plot
the binary periods versus implied velocity for the binary pulsars listed in Table \ref{tab:vels}. This figure should be compared to
the plots in Figure 2 of \citet{tb96}. For system with $P_b$~$<$~2 days and $P_b$~$>$~2 days we find average 2D velocities of
135$\pm$52~km~s$^{-1}$ and 107$\pm$14~km~s$^{-1}$, respectively (uncorrected 2D velocities have averages 
of 120$\pm$45~km~s$^{-1}$ and 85$\pm$11~km~s$^{-1}$). 

We therefore find no significant difference in the velocities
of short- and large-period binary MSPs. However, we caution that only a handful of the former systems are known. 
While the very large velocity of PSR B1957+20 could be explained by these models using asymetries in the supernova 
explosion, they would have a particularly hard time explaining the small implied velocities for PSR J0751+1807 
and PSR 2051$-$0827 given their very short orbital periods. 
In addition, we did simulations using the \citet{tb96} binary period$-$velocity relationship taking into account
the effect of random projections towards us of these velocities. We find that the large scatter in the relationship and
the random projections of these velocities would most likely wash out any effect. Therefore, while we find no evidence that
the relationship is present (and this might indeed be very difficult to achieve, even if it exists), at the same time
we cannot rule it out. It is clear that additional work is needed to understand the evolution of binary MSPs. Obtaining 
additional velocity measurements for these pulsars will help to constrain evolutionary models.

\subsection{Component masses and change in projected semimajor axis}\label{sec:xdot}
For \nineteenten\ and \twenty\ we were able to measure a change in the projected semimajor
axis, $\dot{x}$=$dx/dt$ (see Table \ref{tab:pars}). For \eighteen, the measurement of $\dot{x}$ was marginal 
and will be discussed at the end of this section. For \nineteenfiftythree\ only an upper limit was measured. 
Here we define $x$$\equiv$$a_1$$\sin{i}$/$c$, where $a_1$ is the semimajor 
axis of the pulsar orbit, $i$ is the inclination angle of the angular momentum vector of the orbit relative to the Earth-pulsar line 
of sight (LOS), and $c$ is the speed of light. The measured values for $\dot{x}$ are $-$1.8(5)$\times$10$^{-14}$ and 
8.3(14)$\times$10$^{-14}$ for \nineteenten\ and \twenty, respectively. In principle, a non-zero value for $\dot{x}$ could
arise from a change in $a_1$, $i$ or a combination of the two. However, we argue that the measured values likely arise due 
to the pulsars' high proper motion inducing a change in our LOS to these binaries. 

In the case that a change in $a_1$ is being observed, GR predicts a value of $|\dot{a_1}|$~$\sim$~5$\times$10$^{-23}$ 
and $\sim$10$^{-24}$ for \nineteenten\ and \twenty, respectively \citep[see][for the required expression]{pet64}. These 
values are many orders of magnitude below the observed value of $\dot{x}$. In addition, for typical binary astrophysical 
processes, a non-zero value for  $|$$\dot{a_1}$/$a$$|$ is expected to have a similar order of magnitude as 
$|$$\dot{P_b}$/$P_b$$|$. No significant value for $\dot{P_b}$ was found for any of our pulsars, but allowing for a measurement
in our timing solution results in a value of $\dot{P_b}$=$-$2(4)$\times$10$^{-11}$ and $-$1(2)$\times$10$^{-9}$~s~s$^{-1}$
for \nineteenten\ and \twenty, respectively. Using the values for $P_b$ listed in Table \ref{tab:pars} we find limits of
$|\dot{P_b}/P_b|<$1.2$\times$10$^{-17}$~s$^{-1}$ and $<$5$\times$10$^{-17}$~s$^{-1}$ for \nineteenten\ and \twenty, 
respectively. Again, these values are a few orders of magnitude lower than expected from the measured values of $\dot{x}$.

We therefore propose that the observed values of $\dot{x}$ must arise from apparent changes in the orbital parameters due to the
proper motion of the binaries \citep{ajrt96,kop96}. The strength of this geometric effect can be calculated using:
\begin{equation}\label{eq:xdot}
\dot{x}=1.54\times10^{-16}x\mu\cot{i}\sin{\theta} \;\; \text{s s$^{-1}$}
\end{equation}
where $x$ is the projected semimajor axis in units of seconds, $\mu$ is the total proper motion of the system in units
of mas/yr and $\theta$ is the unknown angle between the position angle of the 
proper motion and the position angle of the ascending node of the pulsar's orbit. The measured values of $\dot{x}$ can 
then be used to constrain the inclination angle of the system $i$. Following \citet{nss01}, we have used a 
Monte Carlo simulation to constrain the values of $i$ that satisfy Equation \ref{eq:xdot} for both \nineteenten\ and
\twenty, resulting in 1$\sigma$ values of 44$^\circ$(36$^\circ$--52$^\circ$) and 36$^\circ$(27$^\circ$--45$^\circ$), 
respectively.

In addition, numerical studies of the evolution of neutron star binaries in long-period orbits with low-mass white dwarf 
companions point to a relationship between the final orbital period, $P_b$, and the mass of the companion, $m_2$ 
\citep{rpj+95,ts99}. An overall agreement with these results has been found in the available data, although these 
relationships appear to overestimate $m_2$ for systems with long periods and provide conflicting results
for systems with short periods \citep{sfl+05}. 

\begin{figure}
\begin{center}
\includegraphics[width=9.5cm]{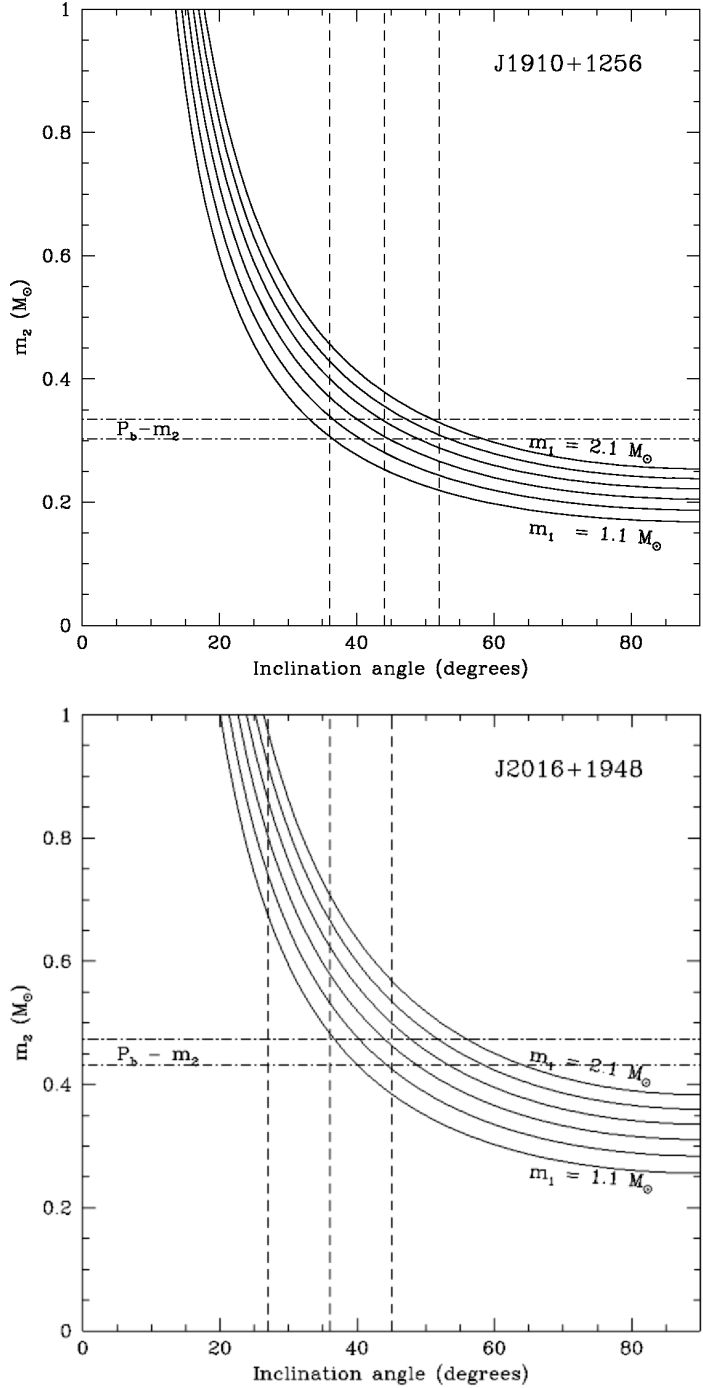}
\caption{\label{fig:m2-i} Constraints on the inclination angle, $i$, and companion mass, $m_2$, for \nineteenten\ ({\it top}) 
and \twenty\ ({\it bottom}). Solid line are constraints derived from the measured mass function of the systems.
Vertical dashed lines represent inclination angle constraints derived from the measured $\dot{x}$ values (centre line is the 
median likelihood and outer lines represent the 1$\sigma$ likelihood limits). Horizontal dot-dashed lines are the $m_2$ values
derived from the $P_b-m_2$ relationship in \citet{ts99}. See \S\ref{sec:xdot} for details.}
\end{center}
\end{figure}

Keeping these caveats in mind, we used the $P_b-m_2$ relationship
found by \citet{ts99}, together with the inclination angle constrains from $\dot{x}$, to provide constraints on 
the masses of the \nineteenten\ and \twenty\ systems. The derived constraints on the companion masses are
$m_2$~=~0.30--0.34~\msun\ and 0.43--0.47~\msun\ for \nineteenten\ and \twenty, respectively. These values and those
implied by the mass functions are shown in 
Figure \ref{fig:m2-i}. Restricting $m_2$ to lie in the values implied by the $P_b-m_2$ relationship and using the inclination
angles from $\dot{x}$ we find 1$\sigma$ values for the pulsar masses of $m_1$~=~1.6$\pm$0.6~\msun\ and 
1.0$\pm$0.5~\msun\ for \nineteenten\ and \twenty, respectively. 

For \eighteen, the measured value of $\dot{x}$ is marginal (see Table \ref{tab:pars}) but can still be used to derive
initial constraints on the system masses. Following the same procedure as outlined above results in a 1$\sigma$ confidence
interval for the inclination angle of this system of 48$^\circ$(33$^\circ$--58$^\circ$). In addition, the  $P_b-m_2$ relationship 
produces companion masses of $m_2$~=~0.33--0.37~\msun. Combining these results, we derive 1$\sigma$ values for
the mass of  \eighteen\ of $m_1$~=~1.4$\pm$0.7~\msun.

The derived $m_1$ values are not very constraining, though fully within the expected mass ranges for neutron stars. 
Given that \twenty\ is only one of three WBMSP with $P_b>$~200 days (see Table \ref{tab:seppsrs}),
further constraining the masses of  this system by independent measurements and continued timing can provide a 
valuable constrain to binary pulsar evolution models.


\subsection{Theories of Gravity: Tests}\label{sec:sep}
We have used the improved timing parameters for the four WBMSPs studied here, in addition to recently discovered
systems, to update important tests of GR and other theories of gravity. In particular, we have modeled the forced eccentricity 
that would be imparted on the binary systems due to violations of the strong equivalence principle (SEP) using the parameter
$\Delta$, and the forced eccentricity that would be imparted due to violations of Lorentz invariance/momentum conservation
using the parameter $\hat{\alpha}_3$. In GR, both parameters are predicted to be identically zero and $\hat{\alpha}_3$ is 
also predicted to be zero in most other theories of gravity.

For the SEP test parameter $\Delta$, 
the additional, forced eccentricity imparted on the binary orbit is expected to be of the form \citep{ds91}:
\begin{equation}\label{eq:efdelta}
|{\bf e}_{F,\Delta}| = \Delta \frac{|{\bf g}_\perp| c^2}{2FGM(2\pi/P_b)^2}
\end{equation}
where $c$ is the speed of light, $F$ is unity in GR and a function of $m_1$ and $m_2$ in alternate theories, $G$ is
Newton's constant in GR, $M$=$m_1$+$m_2$, $P_b$ is the binary period and $|{\bf g}_\perp|$ is the projection of the Galactic
acceleration vector onto the orbital plane at the location of the pulsar. 
Here, the total observed eccentricity is then predicted to
be ${\bf e}_{obs} = {\bf e}_N + {\bf e}_{F,\Delta}$, with the ``natural" eccentricity ${\bf e}_N$ and the angle $\theta$ between ${\bf e}_N$ 
and ${\bf e}_{F,\Delta}$ being additional, unknown parameters.

For the Lorentz invariance/momentum conservation test parameter $\hat{\alpha}_3$, the forced eccentricity added to the
binary orbit is expected to be given by \citep{bd96}:
\begin{equation}\label{eq:efalpha}
|{\bf e}_{F,\hat{\alpha}_3}| =\hat{\alpha}_3 \frac{c_P |\textbf{\emph{V}}|}{24 \pi}\frac{P^2_b}{P}\frac{c^2}{G M} \sin \beta
\end{equation}
where $c_P$=$-$2$E_P^{grav}$/$m_1c^2$ is the gravitation self-energy fraction of the pulsar \citep[the so-called 
``compactness", taken to be approximately 0.21$m_1$;][]{de92,bd96}, 
$\beta$ is the unknown angle between the pulsar system's absolute velocity $\textbf{\emph{V}}$ (with respect to the
reference frame of the cosmic microwave background) and the pulsar's spin vector. 

We have used the above expressions and a Bayesian analysis to derive probability distributions for 
$\Delta$ and $\hat{\alpha}_3$ given the measured binary/pulsar parameters and additional estimates for the remaining
unknown parameters. The procedure was described in detail in \citet{sfl+05} and we summarize it here\footnote{We have
also fixed some small bugs in the \citet{sfl+05} code that did not significantly affect their results.}. 
For each pulsar $j$, we find the probability density functions (pdf) $p(\big| \Delta \big| |D_j,I)$ and 
$p(\hat{\alpha}_3|D_j,I)$ for probable values of $\Delta$ and $\hat{\alpha}_3$ given each pulsar's data 
$D_j$ (see Table \ref{tab:seppsrs}) and prior relevant information $I$.

For example, for the SEP parameter $\Delta$ we can write for each pulsar:
\begin{equation}\label{eq:pdf}
p(\big|\Delta\big|,d_j|D_j,I) \\ \propto p(D_j| \big|\Delta\big|,d_j,I) \times p(\big|\Delta\big|,d_j | I)
\end{equation}
where $d_j$ represents the relevant parameters for this test, namely $i$, $m_2$, $\Omega$, $d$, $e_N$, and $\theta$.
For these parameters we perform Monte Carlo simulations when they are not directly measured or constrained through timing
or other methods. For $m_2$ we use twice the range given by the $P_b-m_2$ relationship of \citet{ts99}. For $\cos i$ 
we assume a uniform distribution between 0 and 1, and combine this value with the measure mass function and 
$m_2$ to provide  a value for the pulsar mass $m_1$ (only systems with $m_1$ values between 1.0 and 2.5 are 
kept)\footnote{For the pulsars presented here with new measured values for $\dot{x}$, we have not included the
implied orbital constaints as they have large error bars.}. For $\Omega$ we use a uniform distribution between 
0$^\circ$ and 360$^\circ$. For $d$ we use a Gaussian
centered on the DM estimate using \cite{cl02b} and assuming an average uncertainty of 25\%, or a Gaussian
centered on the parallax measurement if available. The integrals over the remaining unknown parameters $e_N$ and 
$\theta$ are computed separately using the measured values of ${\bf e}_{obs}$ and $\omega$ and the implied values 
of ${\bf e}_{F,\Delta}$. A pdf for the $\hat{\alpha}_3$ parameter was similarly derived; for this we also need estimates
for the 3D velocity of each pulsar and used Gaussian distributions in each dimension centered on the Galactic 
rotational velocity vector at the pulsar location with widths of 80~km~s$^{-1}$ \citep{lml+98} or, when
available, we use Gaussian distributions for the proper motions to get the transverse velocities. For $|\Delta|$, the
parameter space 10$^{-5}< |\Delta |<$0.1 was sampled uniformly in steps of 2$\times$10$^{-5}$ and for 
$|\hat{\alpha}_3|$ the parameter space 10$^{-22}<|\hat{\alpha}_3|<$5$\times$10$^{-19}$ was sampled
uniformly in steps of 1$\times$10$^{-22}$.

Binary systems suitable for these studies are required to have large periods and small eccentricities so that additional 
relativistic effects are negligible. Large values of $P_b^2/e$ and $P_b^2/Pe$ have therefore been used as a general 
selection characteristic for choosing appropriate systems \citep{wex00,sfl+05}. In addition, the systems must be old 
enough and have large enough 
$\dot{\omega}$ so that the orientation of their orbits can be assumed to be random and that the projection of the Galactic 
acceleration vector on the orbit can be assumed to have been constant over the lifetime of the systems \citep{ds91,wex97}. 
While some pulsars might individually provide low limits for these tests, we use all 27 available systems in order to 
provide a more conservative upper limit that incorporates the assumptions made on the population as a whole. 

\begin{figure}
\begin{center}
\includegraphics[width=15cm]{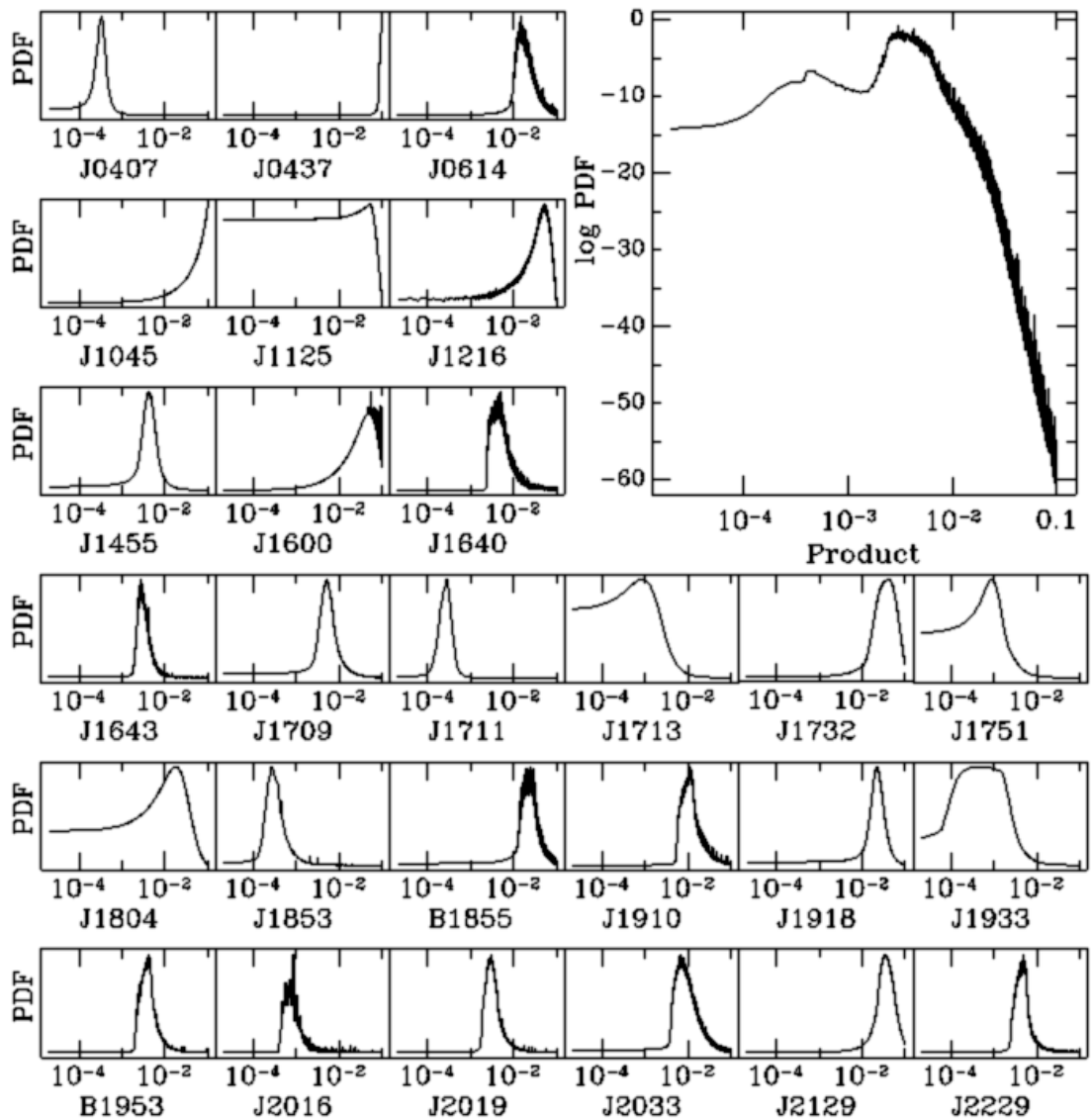}
\caption{\label{fig:sep} Posterior pdf's for the SEP parameter $|\Delta|$. For the individual pulsars, the y-axis is displayed
on a linear scale and the x-axis on a logarithmic scale. The ``product" pdf $p(|\Delta| | D, I)$ in the top centre is the 
normalized product of the pdf's from the individual pulsars and it is shown on a log-log scale.}
\end{center}
\end{figure}

\begin{figure}
\begin{center}
\includegraphics[width=15cm]{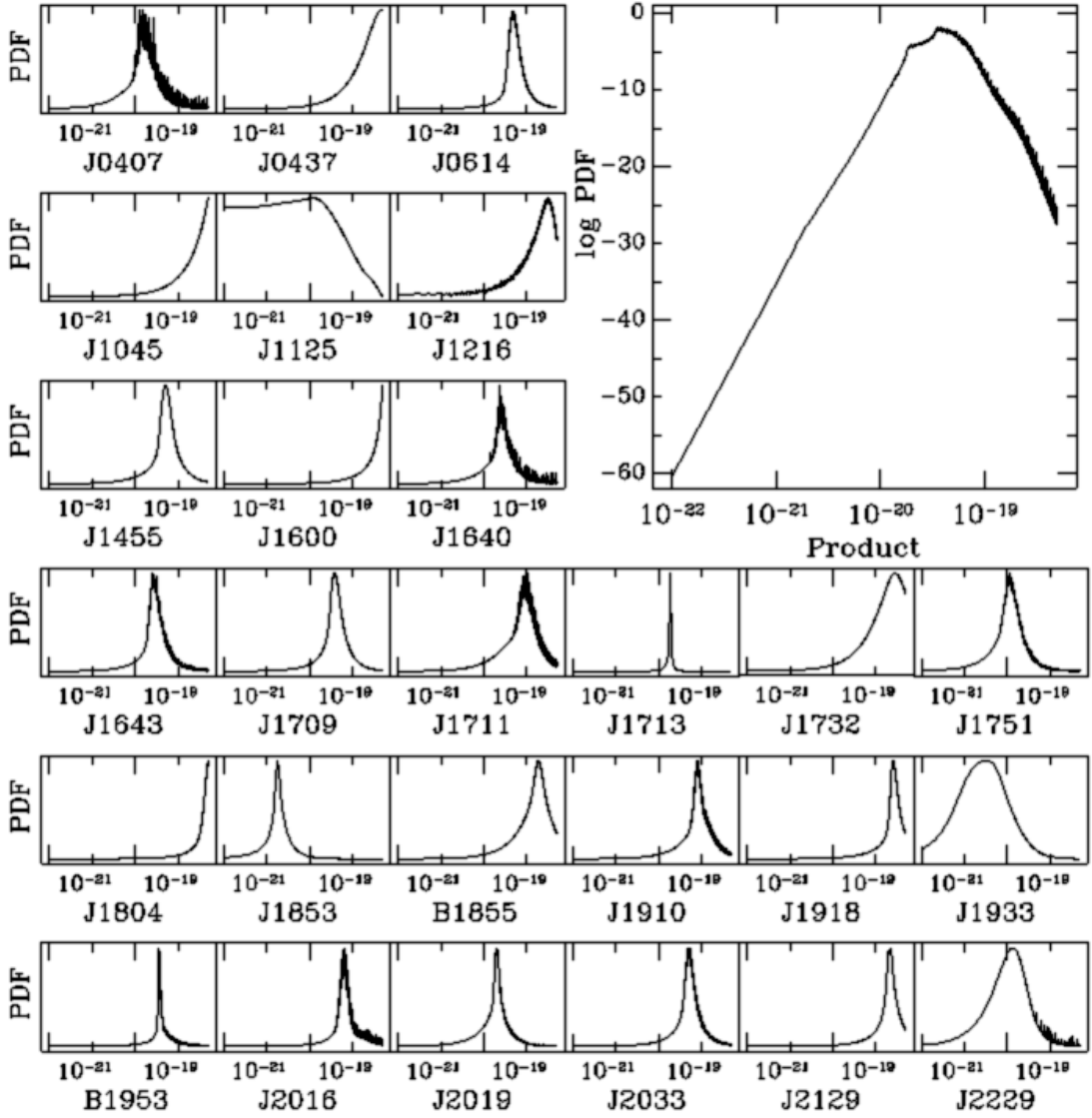}
\caption{\label{fig:a3} Same as Figure \ref{fig:sep} but for the Lorentz invariance/momentum conservation parameter
$|\hat{\alpha}_3|$.}
\end{center}
\end{figure}

Currently, a total of 27 WBMSPs are available to test the above effects and their properties are listed in Table \ref{tab:seppsrs}. 
The pdfs for each pulsar are shown in Figure \ref{fig:sep} for the SEP parameter $|\Delta|$ and in Figure \ref{fig:a3} for the
Lorentz invariance/momentum conservation parameter $|\hat{\alpha}_3|$. Since each pulsar represents an independent
test of these parameters, we can multiply the individual pdfs to obtain a total pdf from our sample of pulsars. 

Using all the systems in Table \ref{tab:seppsrs}, for $|\Delta|$ we derive a 95\% upper limit of 4.6$\times$10$^{-3}$, 
which represents a 20\% improvement from the
value derived by \citet{sfl+05}. Two new pulsars are particularly constraining for this test:
PSR J1711$-$4322 and PSR J1933$-$6211, which together with the improved parameters for \eighteen\ have 
significantly contributed to the reduced upper limit for $|\Delta|$ (the secondary peak at low values of $|\Delta|$ of 
$\sim$10$^{-3.5}$ in the product pdf shown in Figure \ref{fig:sep} is mainly due to these pulsars). For 
PSR J1711$-$4322 alone, the 95\% upper limit for $|\Delta|$ is 5.6$\times$10$^{-4}$.
Since pulsars test gravitational theories in the regime of strong fields, future 
improvements to the above limit are important. The fact that two pulsars discovered in the last five years were able to 
significantly contribute to this test is encouraging and raises the possibility that additional discoveries, and improved 
parameters for the objects already known (particularly PSR J1711$-$4322 and PSR J1933$-$6211), will improve the limit 
further. 

For $|\hat{\alpha}_3|$, using the updated sample of pulsars we derive a 95\% upper limit of 5.5$\times$10$^{-20}$. This limit 
is higher than the value of 4$\times$10$^{-20}$ derived by \citet{sfl+05}. 
We believe that the higher value better reflects
the limits of this technique to constrain $|\hat{\alpha}_3|$ when a larger sample of pulsars is available. The most constraining
pulsars for this test currently are PSR J1713+0747 and PSR J1853+1303 with very similar limits of 2.8$\times$10$^{-20}$ and 
3.1$\times$10$^{-20}$ (95\% confidence), respectively. The limits derived here are about 
13 orders of magnitude smaller than Solar System values \citep{wil93} and again test the strong field limit. Further discoveries 
and ongoing study of present systems (particularly PSR J1853+1303) will help to place additional constraints on this test of 
gravitational theories.

\section{Conclusions}\label{sec:conclusion}
We have presented updated timing solutions for five MSPs, four of which are in binary systems and one which is isolated. 
The high precision and large time span of the observations used allowed us to measure the proper motion
of these pulsars. The implied 2D space velocities in each pulsar's standard of rest lie in the range 70--210~km~s$^{-1}$. 
We studied the available 2D velocities of binary and isolated MSPs and find that their velocity distributions are 
indistinguishable with the current data. 
For \nineteenten\ and \twenty, we are able to measure a significant rate of change
of the semimajor axis which we attribute to a geometrical change in our line of sight to the pulsars due to 
their high space velocities. We are then able to put initial constraints on the mass of these pulsars of 
1.6$\pm$0.6~\msun\ and 1.0$\pm$0.5~\msun\ for \nineteenten\ and \twenty, respectively. For 
\eighteen\ we measured a marginal rate of change of the semimajor axis, resulting in an estimate
for the pulsar mass of 1.4$\pm$0.7~\msun.

We are also
able to place updated constraints on violations of the SEP and Lorentz invariance/momentum 
conservation using an updated list of binary pulsars in wide orbits with small eccentricities. Using a total of 27 pulsars
we derive an upper limit for the SEP violation parameter $|\Delta|$ of 4.6$\times$10$^{-3}$ (95\% confidence)
and an upper limits for the Lorentz invariance/momentum conservation violation parameter $|\hat{\alpha}_3|$ of 
5.5$\times$10$^{-20}$ (95\% confidence). 
Further discoveries and ongoing study of present systems will help to improve these limits.

\acknowledgements
The Arecibo Observatory, a facility of the National Astronomy and Ionosphere Center, is operated by Cornell University 
under a cooperative agreement with the National Science Foundation. The Parkes Radio Telescope is part of the Australia 
Telescope, which is funded by the Commonwealth of Australia for operation as a National Facility managed by CSIRO.
We would like to thank the many collaborators who have helped with this project over the years, including D. Backer, 
D. Lorimer, M. McLaughlin and A. Lommen.  D.J.N. was supported by NSF grant AST 0647820
to Bryn Mawr College. M.E.G. was partly funded by an NSERC PDF award. I.H.S. held an NSERC 
UFA during part of this work, and also acknowledges sabbatical support from the ATNF Distinguished Visitor program and 
from the Swinburne University of Technology Visiting Distinguished Researcher Scheme. Pulsar research at UBC is 
supported by an NSERC Discovery Grant.

\bibliography{journals_apj,modrefs,psrrefs,crossrefs}


\onecolumn

\newpage

\begin{table}[p]
\begin{center}  
\begin{minipage}{1.\textwidth}
\scriptsize
   \centering
\caption{\protect \centering Summary of Observations \label{tab:obs}}
\begin{tabular}{cccccc} 
\hline\noalign{\smallskip}
Telescope & Instrument & Number of & MJD & Center & Effective \\
& & TOAs & Range & Frequencies & Bandwidth\\
& & & & (MHz) & (MHz) \\
\hline\hline\noalign{\smallskip}
\multicolumn{4}{l}{\eighteen:} \\
Arecibo & ASP & 494 & 53370--55105 & 1400 & 64 or 96\\
& & 23 & 54999--55105 & 2350 & 64 or 96\\
& WAPP & 41/32/38 & 53061--55134& 1170/1370/1470 & 50/50/50 \\
& & 7/8/6 & 54882--55134 & 2650/2750/2850 & 100/100/100\\ 
Parkes & Filterbank & 46 &52606--54023& 1390 & 256\\
\hline\noalign{\smallskip}
\multicolumn{4}{l}{\nineteenofive:} \\
Arecibo & ASP & 371 & 53370--54808 & 1400 & 64 or 96 \\
& WAPP & 29/23/39 &  52279--54808 & 1170/1370/1470 & 50/50/50 \\
Parkes & Filterbank & 87 & 51492--53835 & 1390 & 256\\ 
\hline\noalign{\smallskip}
\multicolumn{4}{l}{\nineteenten:} \\
Arecibo & ASP & 430 & 53370--55105 & 1400 & 64 or 96\\
& & 49 &  54882--55105 & 2350 & 64 or 96 \\
& WAPP & 32/26/22 & 53187--55171 & 1170/1370/1470 & 50/50/50 \\
& & 8/9/9 & 54882--55171 & 2650/2750/2850 & 100/100/100 \\
Parkes & Filterbank & 48 & 52602--54062 & 1390 & 256 \\
\hline\noalign{\smallskip}
\multicolumn{4}{l}{\nineteenfiftythree:} \\
Arecibo & ASP & 205 & 53912--55105 & 1400 & 64 or 96\\
& & 15 &  54967--55106 & 2350 & 64 or 96 \\
& WAPP & 17/13/13 & 53912--55134 & 1170/1370/1470 & 50/50/50 \\
& & 2/2/2 & 54882--55171 & 2650/2750/2850 & 100/100/100 \\
& Mark II & 47 & 46112--49096 & 430 & 0.96 \\
\hline\noalign{\smallskip}
\multicolumn{4}{l}{\twenty:} \\
Arecibo & PSPM &  324 & 52456--53591 & 433 & 7.68\\
& WAPP & 775/355/557/199  & 52939--55392 & 1170/1310/1410/1510 & 100/100/100/100 \\
\hline\noalign{\smallskip}
\end{tabular}
\end{minipage}
\end{center}
\end{table}

\begin{sidewaystable}
\begin{minipage}{.85\textwidth}
\scriptsize
   \centering
\caption{\protect \centering Measured and Derived Parameters for the Observed Pulsars \label{tab:pars}}
\begin{tabular}{lccccc} 
\hline\noalign{\smallskip}
Parameter & \eighteen & \nineteenofive & \nineteenten & \nineteenfiftythree & \twenty \\
\hline\noalign{\smallskip}
Right Ascension (RA), $\alpha$ (J2000.0) \dotfill& 18:53:57.319174(8) & 19:05:28.273436(16) & 19:10:09.701479(8) & 19:55:27.87600(3) & 20:16:57.44349(6)\\
Declination (Dec), $\delta$ (J2000.0) \dotfill& 13:03:44.0784(2) & 04:00:10.8830(6) & 12:56:25.5074(3) & 29:08:43.4659(5) & 19:47:51.5882(12)\\
Epoch (MJD) \dotfill& 54000 & 53700 & 54000 & 54500 & 53000.0\\
Data span (MJD) \dotfill & 52606.1--55134.8 & 51492.2--54808.8 & 52602.2--55171.8 & 46112.6--55134.9 & 52456.2--55392.2\\
Proper motion in RA, $\mu_\alpha$=$\dot{\alpha} \cos \delta$ (mas yr$^{-1}$) \dotfill& $-$1.68(7) & $-$3.80(18) & 0.21(10) & $-$0.9(1) & 1.28(26)\\
Proper motion in Dec, $\mu_\delta = \dot{\delta}$ (mas yr$^{-1}$) \dotfill& $-$2.94(12) & $-$7.3(4) & $-$7.25(12) & $-$4.1(1) & 2.83(34)\\
Annual parallax, $\pi$ (mas) \dotfill  & 1.0(6) & $<$2.5$^a$ & $<$0.7$^a$ & $<$7$^a$ & $<$4.5$^a$\\
Spin frequency, $\nu$ (s$^{-1}$) \dotfill& 244.3913778653740(15) & 264.242346143483(16) & 200.658805375034(1) & 163.04791306911(2) & 15.3987376281305(13)\\
Spin frequency derivative, $\dot{\nu}$ (s$^{-2}$) \dotfill & $-$5.2060(5)$\times$10$^{-16}$ & $-$3.425(1)$\times$10$^{-16}$ & $-$3.900(2)$\times$10$^{-16}$& $-$7.901(3)$\times$10$^{-16}$ & $-$9.4997(14)$\times$10$^{-17}$\\
Dispersion measure, DM (pc cm$^{-3}$) \dotfill & 30.5701(6) & 25.6923(12) & 38.0701(8) & 104.501(3) & 33.8148(16)\\
First DM derivative, $\dot{\rm DM}$ (pc cm$^{-3}$ yr$^{-1}$) \dotfill & $<|4\times$10$^{-4}|^a$ &$-$1.1(7)$\times$10$^{-3}$ & 
$< |6\times$10$^{-4}|^a$ & $-$4.7(2.5)$\times$10$^{-3}$$^b$ & $-$1.35(12)$\times$10$^{-3}$$^b$\\
Orbital period, $P_b$ (days) \dotfill& 115.65378643(2) & & 58.466742029(12) & 117.34909728(4) & 635.02377864(7)\\
Project semimajor axis, $x$ (lt-s) \dotfill& 40.7695198(3) & & 21.1291036(3) & 31.4126903(8) & 150.773037(2)\\
Rate of change of $x$, $\dot{x}$ (ls-s s$^{-1}$)\dotfill & 1.7(7)$\times$10$^{-14}$ & & $-$1.8(5)$\times$10$^{-14}$ & $<$4$\times$10$^{-14}$ $^a$ & 8.3(14)$\times$10$^{-14}$\\
Eccentricity, $e$\dotfill & 2.3691(12)$\times$10$^{-5}$ & & 2.3018(3)$\times$10$^{-4}$ & 3.3025(5)$\times$10$^{-4}$ & 1.47981(2)$\times$10$^{-3}$\\
Longitude of periastron, $\omega$ (deg) \dotfill& 346.60(4) & & 106.014(9) & 29.485(8) & 95.6398(7)\\
Epoch of periastron, $T_0$ (MJD) \dotfill& 54046.78(1) & & 54079.318(1) & 54444.267(2) & 52818.648(1)\\
Number of TOAs \dotfill & 695 & 549 & 633 & 316 & 2210\\
Weighted rms residual ($\mu$sec) \dotfill& 1.5 & 5.4 & 1.8 & 3.8 & 11.5\\
\hline\noalign{\smallskip}
\multicolumn{6}{c}{Derived Parameters} \\
\hline\noalign{\smallskip}
Spin period, $P$ (ms)\dotfill & 4.09179744490025(2) & 3.78440484691321(7) & 4.9835840178364(1) & 6.1331666053350(1) & 64.940389248427(5) \\
Spin period derivative, $\dot{P}$ (s s$^{-1}$)\dotfill & 8.7163(8)$\times$10$^{-21}$ & 4.905(2)$\times$10$^{-21}$ & 9.687(4)$\times$10$^{-21}$ & 2.9734(1)$\times$10$^{-20}$ & 4.0063(6)$\times$10$^{-19}$ \\
Surface magnetic field, $B$ (G) \dotfill& 1.9$\times$10$^{8}$ & 1.4$\times$10$^{8}$ & 2.2$\times$10$^{8}$ & 4.3$\times$10$^{8}$ & 5.2$\times$10$^{9}$ \\
Spindown luminosity, $\dot{E}$ (10$^{33}$ ergs s$^{-1}$) \dotfill & 5.1 & 3.5 & 3.1 & 5.1 & 5.8 \\
DM-derived distance, $d_{DM}$ (kpc) \dotfill & 2.1 & 1.7 & 2.3 & 4.64 & 2.5 \\
Characteristic age, $\tau_c$ (Gyr) \dotfill& 7.4 & 12.2 & 8.1 & 3.3 & 2.6 \\
Mass function, $f_1$ (\msun) \dotfill & 0.00543963576(7) & & 0.002962840(12) & 0.00241678837(2) & 0.00912586(4)\\ 
Minimum companion mass$^c$, $m_2$ (\msun)\dotfill & 0.24 & &  0.19 & 0.18 & 0.29 \\
Total proper motion, $\mu$ (mas y$^{-1}$)\dotfill & 3.39(11) & 8.24(36) & 6.98(14) & 4.2(1) & 3.11(32) \\
Galactic angle of proper motion$^d$, $\Theta_\mu$\dotfill & 274$^\circ$& 270$^\circ$ & 210$^\circ$ & 197$^\circ$ & 277$^\circ$ \\
\hline\noalign{\smallskip}
\end{tabular}
\parbox{\textwidth}{Note$-$ Values in parentheses are uncertainties in the last digits shown, which are twice the formal errors quoted
by TEMPO after scaling the TOA uncertainties to obtain $\chi_\nu^2\simeq$1. Right ascension values are in hours, minutes, and 
seconds and declination in degrees, arcminutes, and arcseconds. For all pulsars, the DE405 ephemeris was used and the 
recorded observatory times were corrected to TT(BIPM). \\
$^a$ Value shown represents a $\Delta \chi ^2 \sim$~6.6 from best fit, representing a $\sim$3$\sigma$ limit \citep{avni76}.\\
$^b$ For \twenty, a second DM derivative ($\ddot{DM}$) was also measured with a value of 3.8(1.4)$\times$10$^{-4}$~pc~cm$^{-3}$~yr$^{-2}$. A less significant value for $\ddot{DM}$ was also measured for 
\nineteenfiftythree\ giving $-$2.1(1.6)$\times$10$^{-4}$~pc~cm$^{-3}$~yr$^{-2}$.\\
$^c$ Assuming a pulsar mass of $m_1$ = 1.35\msun.\\
$^d$ Clockwise from Galactic North.}
\end{minipage}
\end{sidewaystable}

\begin{table}[p]
\begin{center}  
\begin{minipage}{1.\textwidth}
\scriptsize
   \centering
\caption{\protect \centering Proper Motions and Space Velocities of Millisecond Pulsars \label{tab:vels}}
\begin{tabular}{lcccrcc} 
\hline\noalign{\smallskip}
Pulsar &  $\mu_\alpha$ & $\mu_\delta$ & Distance & $P_b$ & $V_{2D}$ & Ref.\\
& (mas yr$^{-1}$) & (mas yr$^{-1}$) & (pc) & (days) & (km s$^{-1}$) &\\
\hline\noalign{\smallskip}
\multicolumn{7}{c}{Isolated MSPs} \\
\hline\noalign{\smallskip}
J0030+0451 & $-$5.5$\pm$0.9 & $<-$11 & 310$^a$ &\nodata & $<$20 & 1  \\
J0711$-$6830 & $-$15.55$\pm$0.08 & 14.23$\pm$0.07 & 860 &\nodata &192$\pm$48 & 2 \\
J1024$-$0719 & $-$34.9$\pm$0.4 & $-$47$\pm$1 & 200 &\nodata &48$\pm$13  & 3 \\
J1730$-$2304 & 20.27$\pm$0.06 & \nodata & 510 &\nodata &52$\pm$13 & 4 \\
J1744$-$1134 & 18.804$\pm$0.015 & $-$9.40$\pm$0.06 & 416$^a$ &\nodata &37$\pm$4 & 2 \\
J1905+0400 & $-$3.80$\pm$0.18 & $-$7.3$\pm$0.4 & 1700 &\nodata &89$\pm$25 & This work\\
B1937+21 & $-$0.46$\pm$0.02 & $-$0.66$\pm$0.02 & 3580 &\nodata &107$\pm$31 & 5\\
J1944+0907 & 12.0$\pm$0.7 & $-$18$\pm$3 & 1800 &\nodata &197$\pm$58 & 2 \\
J2124$-$3358 & $-$14.12$\pm$0.13 & $-$50.34$\pm$0.25 & 322$^a$ &\nodata &87$\pm$35 & 6\\
J2322+2057 & $-$17$\pm$2 & $-$18$\pm$3 & 790 &\nodata &100$\pm$30  & 7\\
\hline\noalign{\smallskip}
\multicolumn{7}{c}{Binary MSPs} \\
\hline\noalign{\smallskip}
J0437$-$4715 & 121.438$\pm$0.003 & $-$71.438$\pm$0.007 & 157$^a$ & 5.741 & 146$\pm$3 & 8 \\
J0613$-$0200 & 1.84$\pm$0.08 & $-$10.6$\pm$0.2 & 1700 & 1.198 & 100$\pm$23 & 2 \\
J0751+1807 & $-$1.3$\pm$0.2 & $-$6.0$\pm$1.8 & 610 & 0.263 & 18$\pm$4 & 9 \\
J1012+5307 & 2.4$\pm$0.2 & $-$25.2$\pm$0.2 & 840 & 0.604 & 110$\pm$25 & 10\\
J1023+0038 & 10$\pm$1 & $-$16$\pm$2 & 1300 & 0.198 & 90$\pm$26 & 11 \\
J1045$-$4509 & $-$6.0$\pm$0.2 & 5.3$\pm$0.2 & 1940 & 4.083 & 155$\pm$36 & 2\\
J1455$-$3330 & 5$\pm$6 & 24$\pm$12 & 530 & 76.174 & 57$\pm$30 & 4 \\
J1600$-$3053 & $-$0.99$\pm$0.10 & $-$6.7$\pm$0.5 & 2930 & 14.348 & 92$\pm$26 & 12\\
J1640+2224 & 1.66$\pm$0.12 & $-$11.3$\pm$0.2 & 1160 & 175.461 & 69$\pm$18 & 13 \\
J1643$-$1224 & 6.0$\pm$0.1 & 4.1$\pm$0.4 & 454$^a$ & 147.017 & 22$\pm$2 & 2 \\
J1709+2313 & $-$3.2$\pm$0.7 & $-$9.7$\pm$0.9 & 1400 & 22.711 & 83$\pm$23 & 14 \\
J1713+0747 & 4.917$\pm$0.004& $-$3.933$\pm$0.010 & 1050$^a$ & 67.825 & 29$\pm$3 & 15\\
J1853+1303 & $-$1.67$\pm$0.08 & $-$2.91$\pm$0.12 & 2100 & 115.654 & 72$\pm$22 & This work\\
B1855+09 & $-$2.899$\pm$0.013 & $-$5.45$\pm$0.02 & 910 & 12.327 & 37$\pm$10 & 16 \\
J1903+0327 & $-$2.01$\pm$0.07 & $-$5.20$\pm$0.12 & 6400 & 95.714 & 180$\pm$5 & 17\\
J1909$-$3744 & $-$9.510$\pm$0.007 & $-$35.859$\pm$0.019 & 1270$^a$ & 1.533 & 231$\pm$13 & 2\\
J1910+1256 & 0.21$\pm$0.10 & $-$7.25$\pm$0.12 & 2300 & 58.467 & 117$\pm$34 & This work\\
J1911$-$1114 & $-$6$\pm$4 & $-$23$\pm$13 & 1220 & 2.716 & 129$\pm$75 & 4\\
J1918$-$0642 & $-$7.2$\pm$0.1 & $-$5.7$\pm$0.3 & 1240 &  10.913 & 49$\pm$11 & 18\\
B1953+29 & $-$0.9$\pm$0.1 & $-$4.1$\pm$0.1 & 4640 & 117.349 & 214$\pm$56 & This work \\
B1957+20 & $-$16.0$\pm$0.5 & $-$25.8$\pm$0.6 & 2490 & 0.382 & 422$\pm$110 & 19\\
J2019+2425 & $-$9.41$\pm$0.12 & $-$20.60$\pm$0.15 & 1490 & 76.512 & 192$\pm$51 & 20\\
J2033+1734 & $-$5.94$\pm$0.17 & $-$11.0$\pm$0.3 & 2000 & 56.308 & 162$\pm$43 & 16 \\
J2051$-$0827 & 5.3$\pm$1 & 0.3$\pm$3 & 1040 & 0.099 & 20$\pm$10 & 21 \\
J2129$-$5721 & 9.35$\pm$0.1 & $-$9.47$\pm$0.1 & 1360 & 6.625 & 104$\pm$30 & 2\\
J2229+2643 & 1$\pm$4 & $-$17$\pm$4 & 1450 & 93.016 & 126$\pm$45 & 22\\
J2317+1439 & $-$1.7$\pm$1.5 & 7.4$\pm$3.1 & 820 & 2.459 & 40$\pm$15 & 23 \\
 \hline\noalign{\smallskip}
\end{tabular}
\parbox{\textwidth}{References: (1) \citet{lkn+06}, (2) \citet{vbc+09}; (3) \citet{hbo06}; (4) \citet{tsb+99}; (5) \citet{clm+05}; 
(6) \citet{nt95}; (7) \citet{nt95}; (8) \citet{vbv+08}; (9) \citet{nss+05}; (10) \citet{lcw+01}; (11) \citet{asr+09}; (12) \citet{ojhb06}; 
(13) \citet{llww05}; (14) \citet{lwf+04}; (15) \citet{sns+05}; (16) \citet{spl04}; (17) \citet{fbw+11}; (18) \citet{jsb+10}; 
(19) \citet{antt94}; (20) \citet{nss01}; (21) \citet{dlk+01}; (22) \citet{wdk+00}; (23) \citet{cnst96}.\\
$^a$ Distance derived using parallax. See reference for values.}
\end{minipage}
\end{center}
\end{table}

\begin{sidewaystable}
\begin{minipage}{1.\textwidth}
\scriptsize
   \centering
\caption{\protect \centering WBMSPs used for SEP and Lorentz invariance/momentum conservation tests \label{tab:seppsrs}}
\begin{tabular}{lrrccccccc} 
\hline\noalign{\smallskip}
Pulsar& $P$ & $P_b$ & $e$ & $\omega$ & $f_1$ & $\mu_\alpha$ & $\mu_\delta$ &$d$ & Ref. \\
 & (ms) & (days) & & ($^{\circ}$)& (\msun)& (mas yr$^{-1}$) & (mas yr$^{-1}$)  & (kpc) &\\ 
\hline\noalign{\smallskip}
J0407+1607 & 25.7017 & 669.0704 & 9.368(6)$\times$10$^{-4}$ & 291.74(2) & 0.002893 & \nodata &\nodata &1.33 & 1\\
J0437$-$4715 & 5.7574 & 5.7410 & 1.91685(5)$\times$10$^{-5}$ & 1.22(5) & 0.001243 & 121.438$\pm$0.003 &  $-$71.438$\pm$0.007 & 0.157 & 2, 3, 4\\
J0614$-$3329 & 3.1487 & 53.5846 & 1.801(1)$\times$10$^{-4}$ & 15.92(4) & 0.007895 & \nodata & \nodata & 1.9 & 5 \\
J1045$-$4509 & 7.4742 & 4.0835 & 2.37(7)$\times$10$^{-5}$ & 243(2) & 0.001765 &  $-$6.0$\pm$0.2 & 5.3$\pm$0.2 & 1.94 & 6 \\
J1125$-$6014 & 2.6304 & 8.7526 & 1(13)$\times$10$^{-6}$ & 273(87) & 0.008128 &\nodata &\nodata & 1.50 & 7 \\
J1216$-$6410 & 3.5394 & 4.0367 & 7(59)$\times$10$^{-6}$ & 177(1) & 0.001669 &\nodata &\nodata & 1.33 & 7 \\ 
J1455$-$3330 & 7.9872 & 76.1745 & 1.697(3)$\times$10$^{-4}$ & 223.81(1) & 0.006272 & 5$\pm$6 & 24$\pm$12 & 0.53 &  8, 9  \\
J1600$-$3053 & 3.5979 & 14.3484 & 1.7373(2)$\times$10$^{-4}$ & 181.73(1) & 0.003558 & -0.99$\pm$0.10 & $-$6.7$\pm$0.5 & 2.93 & 10 \\
J1640+2224 & 3.1633 & 175.4607 & 7.97262(14)$\times$10$^{-4}$ & 50.7308(10) & 0.005907 & 1.66$\pm$0.12 & $-$11.3$\pm$0.2 &1.16 & 11, 12 \\
J1643$-$1224 & 4.6216 & 147.0174 & 5.0579(4)$\times$10$^{-4}$ & 321.850(4) & 0.000783 & 6.0$\pm$0.1 & 4.1$\pm$0.4 & 0.454 & 6 \\
J1709+2313 & 4.6312 & 22.7119 & 1.87(2)$\times$10$^{-5}$ & 24.3(6) & 0.007438  & $-$3.2$\pm$0.7 & $-$9.7$\pm$0.9 & 1.4 & 11, 13 \\
J1711$-$4322 & 102.6183 & 922.4707 & 2.375(6)$\times$10$^{-3}$ & 293.75(12) & 0.003434 & \nodata &\nodata & 3.84 & 7\\
J1713+0747 & 4.5701 & 67.8251 & 7.49406(13)$\times$10$^{-5}$ & 176.1915(10) & 0.007896 & 4.917$\pm$0.004& $-$3.933$\pm$0.010 & 1.05 & 14, 15\\
J1732$-$5049 & 5.3125 & 5.2630 & 9.8(20)$\times$10$^{-6}$ & 287(12) & 0.002449 & \nodata &\nodata & 1.41 & 16 \\
J1751$-$2857 & 3.9149 & 110.7465 & 1.283(5)$\times$10$^{-4}$ & 45.52(19) & 0.003013 & \nodata &\nodata &  1.1 & 17 \\
J1804$-$2717 & 9.3430 & 11.1287 & 3.1(5)$\times$10$^{-5}$ & 160(4) & 0.003347 & \nodata &\nodata & 0.78 & 18, 19\\
J1853$+$1303 & 4.0918 & 115.6538 & 2.3686(12)$\times$10$^{-5}$ & 346.61(4) & 0.005439 & $-$1.67$\pm$0.08 & $-$2.91$\pm$0.12 &2.1 & This work \\
B1855+09 & 5.3621 & 12.3272 & 2.170(3)$\times$10$^{-5}$ & 276.39(4) & 0.005557 & $-$2.899$\pm$0.013 & $-$5.45$\pm$0.02 & 0.91 & 20, 21 \\
J1910+1256 & 4.9836 & 58.4667 & 2.3017(3)$\times$10$^{-4}$ & 106.016(9) & 0.002962 & 0.21$\pm$0.10 & $-$7.25$\pm$0.12 & 2.3 & This work \\
J1918$-$0642 & 7.6459 & 10.9132 & 1.991(13)$\times$10$^{-5}$ & 218.6(4) & 0.005249 & $-$7.2$\pm$0.1 & $-$5.7$\pm$0.3 & 1.24 & 22 \\
J1933$-$6211 & 3.5434 & 12.8194 & 1.3(4)$\times$10$^{-6}$  & 116(22) & 0.011749 & \nodata &\nodata & 0.52 & 23 \\
B1953+29 & 6.1332 & 117.3491 & 3.3026(5)$\times$10$^{-4}$ & 29.489(8) & 0.002417 & $-$0.9$\pm$0.1 & $-$4.1$\pm$0.1 & 4.64 & This work \\
J2016+1948 & 64.9404 & 635.0238 & 1.47980(2)$\times$10$^{-3}$ & 95.640(1) & 0.009126 & 1.28$\pm$0.26  & 2.82$\pm$0.34&  2.5 & This work \\
J2019$-$5721 & 3.9345 & 76.5116 & 1.1109(4)$\times$10$^{-4}$ & 159.03(2) & 0.010687 &$-$9.41$\pm$0.12 & $-$20.60$\pm$0.15& 1.49 & 24, 25 \\
J2033+1734 & 5.9489 & 56.308 & 1.2876(6)$\times$10$^{-4}$ & 78.23(3) & 0.002776 &  $-$5.94$\pm$0.17 & $-$11.0$\pm$0.3 & 2.0 & 21 \\
J2129$-$5721 & 3.7263 & 6.6255 & 1.21(3)$\times$10$^{-5}$ & 196.3(1.5) & 0.001049 &  9.35$\pm$0.1 & $-$9.47$\pm$0.1 &1.36 & 6 \\
J2229+2643 & 2.9778 & 93.0159 & 2.556(2)$\times$10$^{-4}$ & 14.42(0.05) & 0.000839 & 1$\pm$4 & $-$17$\pm$4 & 1.45 & 26, 27 \\
 \hline\noalign{\smallskip}
\end{tabular}
\parbox{\textwidth}{References: (1) \citet{lxf+05}; (2) \citet{jlh+93}; (3) \citet{vbb+01}; (4) \citet{vbv+08}; (5) \citet{rrc+11}; (6) \citet{vbc+09}; 
(7) \citet{lfl+06}; (8) \citet{tsb+99}; (9) \citet{lnl+95}; (10) \citet{ojhb06}; (11) \citet{fcwa95}; (12) \citet{llww05}; (13) \citet{lwf+04}; 
(14) \citet{fwc93}; (15) \citet{sns+05}; (16) \citet{eb01}; (17) \citet{sfl+05}; (18) \citet{llb+96}; (19) \citet{hllk05}; (20) \citet{srs+86}; 
(21) \citet{spl04}; (22) \citet{jsb+10}; (23) \citet{jbo+07}; (24) \citet{ntf93}; (25) \citet{nss01}; (26) \citet{cnst96}; (27) \citet{wdk+00}.}
\end{minipage}
\end{sidewaystable}

\end{document}